\documentclass[a4paper,11pt]{article}
\usepackage{amsmath,amssymb,graphicx,bezier,citesort}
\newcommand{\nc}{\newcommand}
\nc{\rnc}{\renewcommand}
\nc{\nn}{\nonumber}
\nc{\ch}{\cosh}
\nc{\sh}{\sinh}
\def\i{{\rm i}}
\def\e{{\rm e}}
\rnc{\Im}{{\rm{Im}\,}}
\rnc{\Re}{{\rm{Re}\,}}
\nc{\mfa}{{\mathfrak{a}}}
\nc{\mfab}{\overline{\mfa}}
\nc{\mfA}{{\mathfrak{A}}}
\nc{\mfAb}{\overline{\mfA}}
\nc{\mfb}{{\mathfrak{b}}}
\nc{\mfbb}{\overline{\mfb}}
\nc{\mfB}{{\mathfrak{B}}}
\nc{\mfBb}{\overline{\mfB}}
\nc{\mfsl}{{\mathfrak{sl}}}
\nc{\db}{\displaybreak[0]\\}
\nc{\bra}{\langle}
\nc{\ket}{\rangle}
\nc{\R}{\check{R}}
\nc{\D}{D_{\rm th}}
\nc{\tr}{{\rm Tr}}
\nc{\J}{\mathcal{J}}
\nc{\E}{E^{h}}
\nc{\p}{\frac{\pi}{\gamma}}
\rnc{\(}{\left(}
\rnc{\)}{\right)}
\nc{\K}{\kappa}

\numberwithin{equation}{section}

\begin{document}
%
\title{Non-dissipative thermal transport
in the massive regimes of the $XXZ$ chain}
\author{
Kazumitsu Sakai\thanks{E-mail address:
kaz@issp.u-tokyo.ac.jp}\\\\
\it Institute for solid state physics, 
University of Tokyo, \\
\it Kashiwa, Chiba 277--8581, Japan \\\\
Andreas Kl\"umper\thanks{E-mail address:
kluemper@physik.uni-wuppertal.de} \\\\
\it Bergische Universit\"at Wuppertal, Fachbereich Physik, \\ 
\it D-42097 Wuppertal, Germany 
} 
\date{July 10, 2003} 
\maketitle
%
%
\begin{abstract}
We present 
exact results on the thermal conductivity of the one-dimensional spin-1/2
$XXZ$ model in the massive antiferromagnetic and ferromagnetic regimes. The
thermal Drude weight is calculated by a lattice path integral formulation.
Numerical results for wide ranges of temperature and anisotropy as well as
analytical results in the low and high temperature limits are presented.  At
finite temperature, the thermal Drude weight is finite and hence there is
non-dissipative thermal transport even in the massive regime.
At low temperature, the thermal Drude weight behaves as $D(T)\sim
\exp(-\delta/T)/\sqrt{T}$ where $\delta$ is the one-spinon (respectively
one-magnon) excitation energy for the antiferromagnetic (respectively
ferromagnetic) regime.  
\end{abstract}

%
\section{Introduction}
%
Recently, transport properties of low-dimensional strongly correlated quantum
systems have been extensively studied from both theoretical and experimental
sides (see for example \cite{ZP03} and reference therein).
Among them anomalously enhanced thermal conductivity
\cite{Kudo,Kudo2,Sologubenko00,Sologubenko01,Hess01,Sologubenko03} 
and ballistic spin transport \cite{Taki,Thurber} have been reported 
in experiments on one- or quasi one-dimensional materials with weak 
interchain interactions.
Theoretically, the existence of such anomalous properties has also been
pointed out especially in quantum integrable systems.
One of the criteria for anomalous transport is the existence of a non-zero
Drude weight.

In particular for the spin-1/2 $XXZ$ chain, the spin transport was
investigated analytically \cite{Zotos99} by use of a method developed in
\cite{FK98}. In the massless 
regime, the Drude weight is non-zero at arbitrary temperatures resulting into
non-dissipative (ballistic) spin transport.
On the contrary, in the massive regime without magnetic field, results on the
basis of \cite{Zotos99} and \cite{FK98} indicate a zero Drude weight for any
temperature implying that the spin transport is dissipative.
The validity, however, of these results is still controversial due to the
complexity of the analysis at finite temperatures and debated in
\cite{NarMA98,AlGros1,FK03}.
For the heat conduction, the {\it thermal} Drude weight in the critical regime
was more recently calculated by the Bethe ansatz technique in \cite{KS} which
shows that the thermal transport is non-dissipative at any finite
temperature.  
The study was extended by field theoretical and numerical
approaches to more general models including non-integrable systems, and the
existence of a nonzero Drude weight was actively discussed
\cite{AlGros2,Saito1,Saito2,Saito3,MHCB,MHCB2,OCC03}.

In this paper we discuss the thermal conductivity in the massive regime of the
$XXZ$ model by use of the approach developed in our previous work \cite{KS}.
In the case that the response of a physical system to a perturbation is
proportional to the force, the quantitative relation is obtained from linear
response theory.
Transport coefficients are universally given by the Kubo formulae
\cite{Kubo,Mahan} in terms of correlation functions in thermal equilibrium
without perturbation.
In this way the thermal conductivity $\kappa(\omega)$ relating the thermal
current ${\mathcal J}_{\rm th}$ to the temperature gradient, ${\mathcal
J}_{\rm th}=\kappa \nabla T$, is given by the correlation function of the
thermal current operator $\J_{\rm th}$:
\begin{align}
\kappa(\omega)=\frac{1}{T}\int_0^\infty {\rm d} t \e^{-\i\omega t} \phi(t),
\qquad
\phi(t)= \int_0^{\beta} {\rm d}\tau  
         \langle\J_{\rm th}(-t-\i\tau)\J_{\rm th}\rangle 
\label{kubo}
\end{align}
where $\beta$ is the reciprocal temperature; $\beta=1/T$ and
$\langle\cdots\rangle$ denotes the thermal expectation value per site.
Note that here we do not take into account thermomagnetic effects (cf. Sec.5).
The real part of eq.\eqref{kubo} reduces to
\begin{equation}
\Re \kappa(\omega)= \pi D_{\rm th}(T)\delta(\omega)+
                    \kappa_{\rm reg}(\omega)
\end{equation}
where $\D(T)$ is the thermal Drude weight\footnote{In the previous work
\cite{KS} the thermal Drude weight (written by $\tilde{\kappa}$) is defined as
$\tilde{\kappa}=\pi \D(T)$.} given by
\begin{equation}
\D(T)=\frac{1}{TL}\left\{ \phi(0)-2\sum_{ E_n\ne E_m}p_n 
\frac{\bigl|\langle n |\J_{\rm th} |m\rangle \bigr|^2}
      {E_m-E_n}\right\} 
\quad 
p_n=\frac{\e^{-\beta E_n}}{\sum_k \e^{-\beta E_k}}.
\label{drude1}
\end{equation}

In general, the evaluation of the above correlation function is very
difficult.
Fortunately, as already pointed out by Zotos et al. \cite{znp}, the thermal
current operator of certain integrable systems commutes with the Hamiltonian.
Indeed, the thermal current can be identified as one of the infinitely many
conserved quantities underlying the integrability.
In this fortuitous situation the thermal current correlations are not time
dependent, they reduce to ordinary static correlations.  Hence we find
\begin{equation}
\D(T)=\beta^2\langle \J_{\rm th}^2\rangle.
\label{drude}
\end{equation}
Consequently, this quantity is exactly calculable within a lattice path
integral formulation as described in \cite{KS}. At any finite temperature, the
thermal Drude weight is finite implying non-dissipative thermal transport even
in the massive regimes.  At low temperature, due to the energy gap of the
elementary excitations, the thermal Drude weight decays exponentially with
decrease of temperature (cf. eq.(4.11) in \cite{KS} for the massless regime).

This paper is organized as follows.  In Sec.2 we briefly review the relation
between the thermal current operator and conserved quantities.
In Sec.3 we consider the thermal Drude weight at finite temperatures and
discuss the results. In Sec.4 analytical results in the low- and
high-temperature limits are presented. Sec.5 is devoted to a summary of the
present work and an outlook on the case of finite magnetic field.
%
\section{Ideal thermal current}
%
Let us consider the $XXZ$ model on a periodic chain with $L$ sites:
\begin{align}
\mathcal{H}=\sum_{k=1}^{L}h_{k k+1} \qquad
h_{kk+1}=J\left\{\sigma^{+}_{k}\sigma^{-}_{k+1}+
                        \sigma^{+}_{k+1}\sigma^{-}_{k}+
               \frac{\Delta}{2}
               \left(\sigma^{z}_{k}\sigma^{z}_{k+1}-1\right)
                 \right\}
\label{hami}
\end{align}
where $\sigma_{k}^{\pm}=(\sigma^x_k\pm \i \sigma^y_k)/2$ and $\sigma^x_k,
\sigma^y_k,\sigma^z_k$ denote the Pauli matrices acting on the $k$th space.
The transfer integral $J$ together with the anisotropy parameter $\Delta$
determine the physical nature of the system.
In this paper we focus our attention on the massive regime $\Delta>1$, where
correlation functions decay exponentially at zero temperature.
Here $J>0$ (respectively $J<0$) corresponds to the antiferromagnetic
(respectively ferromagnetic) model.
In this regime it is convenient to introduce the parameter $\gamma$ instead of
$\Delta$:
\begin{equation}
\Delta=\ch\gamma \qquad \gamma\in\mathbb{R}_{\ge 0}.
\end{equation}

Our first aim is to determine the local energy current operator $j_k^{\rm
E}$. To achieve this we relate the time derivative of the local Hamiltonian to
the (discrete) divergence of the thermal current via the continuity equation
$\dot{h}=-{\rm div\,} j^{\rm E}$. Since $\dot{h}$ is written as the commutator
with the Hamiltonian, we obtain
\begin{equation}
\dot{h}_{kk+1}=\i[\mathcal{H},h_{kk+1}(t)]=
-\{j_{k+1}^{\rm E}(t)-j_k^{\rm E}(t)\}.
\label{continuity}
\end{equation}
Obviously the local energy current $j^{\rm E}_k$ defined by
\begin{equation}
j_k^{\rm E}=\i[h_{k-1k},h_{kk+1}]
\label{lc}
\end{equation}
satisfies the last relation in \eqref{continuity}.
For zero magnetic field the energy current operator $\mathcal{J}_{\rm
E}=\sum_{k=1}^{L}j_k^{\rm E}$ is equivalent to the thermal current operator
$\J_{\rm E} =\J_{\rm th}$ (cf. Sec.5 for non-zero magnetic field).
Explicitly it reads
\begin{equation}
\J_{\rm th}=-\i J^2\sum_{k=1}^{L}
    \left\{\sigma_k^{z}(\sigma_{k-1}^{+}\sigma_{k+1}^{-}-
           \sigma_{k+1}^{+}\sigma_{k-1}^{-})-
           \Delta(\sigma_{k-1}^{z}+\sigma_{k+2}^{z})
                 (\sigma_{k}^{+}\sigma_{k+1}^{-}-
                  \sigma_{k+1}^{+}\sigma_{k}^{-})\right\}.
\label{tc}
\end{equation}
As already shown by Zotos et.al in \cite{znp}, this thermal current operator
is a conserved quantity.

To show this from the underlying integrability, we consider the six vertex
model which is the classical counterpart of the $XXZ$ chain.  There are six
spin configurations carrying non-zero Boltzmann weights. This corresponds to
six non-zero elements of the $R$-matrix:
\begin{equation}
R_{11}^{11}(v)=R_{22}^{22}(v)=1\quad 
R_{12}^{12}(v)=R_{21}^{21}(v)=\frac{[v]}{[v+2]}\quad
R_{12}^{21}(v)=R_{21}^{12}(v)=\frac{[2]}{[v+2]}
\end{equation}
where $[v]$ is an abbreviation for $[v]=\sh(\gamma v/2)$ and the meaning of
the indices is exactly as in ref.\cite{KS}.
The original quantum spin chain is connected with the classical model by a
relation of the Hamiltonian $H$ and the row-to-row transfer matrix $T(v)={\rm
Tr}_a\prod_{k=1}^{L}R_{a k}(v)$
\begin{equation}
H=A\frac{{\rm d}}{{\rm d} v}\ln T(v)\Big|_{v=0} \qquad
A=\frac{2 J \sh \gamma}{\gamma}
\label{defH}
\end{equation}
where $T(v)$ is a commuting family with respect to different parameters:
$[T(v),T(v^{\prime})]=0$.
Due to the commutativity, the transfer matrix is a generator of conserved
currents $\mathcal{J}^{(n)}$:
\begin{equation}
\mathcal{J}^{(n)}=\i^{n-1} (A D)^n
\ln T(v) \Big|_{v=0}\quad D=\frac{{\rm d}}{{\rm d} v}.
\label{generator}
\end{equation}
Note that $\mathcal{J}^{\rm (1)}$ corresponds to the Hamiltonian \eqref{hami}.
For $n=2$, we directly find
\begin{equation}
\mathcal{J}^{(2)}=\i A^2 \sum_{k=1}^{L}
\left\{\R''_{k k+1}(0)-\R^{\prime 2}_{k k+1}(0)+
[\R^{\prime}_{k-1k}(0),\R^{\prime}_{kk+1}(0)]
\right\}\quad  \check{R}(v):=P R(v)
\end{equation}
where $P$ denotes the permutation operator.  Thanks to the unitarity
$\R(v)\R(-v)=1$, the first two terms in the above equation cancel!
Using the identity $A \check{R}^\prime_{kk+1}(0)=h_{kk+1}$ and eq.\eqref{lc},
we see that $\mathcal{J}^{(2)}$ coincides with the thermal current $\J_{\rm
th}$.
Due to this and the fact that the thermomagnetic power is always zero for zero
magnetic field (being equivalent to  half-filling), the thermal Drude weight
$\D(T)$ is given by \eqref{drude} (cf. \eqref{drudeh} in Sec.5).

Here we only used the difference property and the unitarity of the $R$-matrix
to show the conservation law of the thermal current operator.
These properties hold not only for the $XXZ$ chain but for any integrable
system whose Hamiltonian is defined as in \eqref{defH}, hence the thermal
current operator for such an integrable system also satisfies the conservation
law\footnote{One of the exceptions is the Hubbard model for which the
$R$-matrix does not have the difference property.}.
%
\section{Thermal conductivity}
%
In order to obtain the thermal Drude weight \eqref{drude}, we have to evaluate
the expectation value of the square of the thermal current operator. In fact,
we are able to derive explicit results for the generating function of the
expectation values of any power of 
any conserved current.
Let us introduce the following extended Hamiltonian $\widetilde{\mathcal{H}}$
including the conserved currents \eqref{generator} as a perturbation;
$\widetilde{\mathcal{H}}:=\mathcal{H}-T \lambda_n \J^{(n)}$ (throughout this
paper we set $\lambda_n\ll 1$).
Introducing the partition function $Z(\lambda_n )=\tr\,\e^{-\beta\widetilde{\mathcal{H}}}$,
we easily see that the autocorrelations of the conserved quantities can be
calculated by taking the second logarithmic derivative with respect to the
variable $\lambda_n $:
\begin{equation}
\langle\J^{(n) 2}\rangle-
\langle\J^{(n)}\rangle^2=\frac{1}{L}\frac{\partial^2\ln Z(\lambda_n )}
{\partial \lambda_n ^2}\bigg|_{\lambda_n =0}.
\label{currentcorr}
\end{equation}
To evaluate $Z(\lambda_n)$ explicitly, we follow a procedure developed in the
previous work \cite{KS}. Taking into account the relation \eqref{generator},
we express the partition function $Z(\lambda_n )$ 
in terms of the row-to-row transfer matrices $T(v)$.
\begin{equation}
Z(\lambda_n )=\lim_{N\to\infty}
\tr \exp\left[T(0)^{-N}\prod_{j=1}^{N}T(u_j)\right]
=\tr \exp
\left[
\lim_{N\to\infty}\sum_{j=1}^{N} \left\{\ln T(u_j)-\ln T(0)\right\}\right] 
\end{equation}
where a sequence of $N$ numbers $u_1,\dots,u_N$ should be chosen such that
\begin{equation}
\lim_{N\to\infty}\sum_{j=1}^{N} \left\{\ln T(u_j)-\ln T(0)\right\}
=A D \left\{-\beta  +\lambda_n  \i^{n-1}(AD)^{n-1}\right\}
\ln T(v)\big|_{v=0}.
\label{general}
\end{equation}

Applying a lattice path integral formulation, we introduce the quantum
transfer matrix (QTM) in the imaginary time direction
\cite{MSuzPB,InSuz,KlumTH,KlumTH2,destri}.
In this formalism the partition function $Z(\lambda_n)$ and the quantity
\eqref{currentcorr} in the thermodynamic limit $L\to\infty$ can be expressed
as the largest eigenvalue of the QTM $\Lambda$:
\begin{equation}
\lim_{L\to\infty}\frac{1}{L}\ln Z(\lambda_n )=\ln \Lambda 
\qquad
\bra \J^{(n)2}\ket-\bra \J^{(n)}\ket^2=
\left(\frac{\partial}{\partial \lambda_n}\right)^2\ln\Lambda.
\label{qtm}
\end{equation}
The integral expression for $\Lambda$ is given by\footnote{For convenience, we
change here the spectral parameter $v\to\i v$.}
\begin{align}
&\ln\Lambda=\left\{-\beta+\lambda_n  (AD)^{n-1}\right\}
\mathcal{E}(v)\big|_{v=0}+
\int_{-\frac{\pi}{\gamma}}^{\frac{\pi}{\gamma}}K(v)
\ln[\mfA(v)\mfAb(v)]{\rm d} v \nn \db
& 
K(v)=\frac{\gamma}{2\pi}
\sum_{k=-\infty}^{\infty}\frac{\e^{-\i k \gamma v}}
{{2\ch k \gamma}}
\label{eigval}
\end{align}
where $\mathcal{E}(0)$ is the ground state energy of the antiferromagnetic
system.  Explicitly $\mathcal{E}(v)$ reads
\begin{equation}
\mathcal{E}(v)=\int_{-\frac{\pi}{\gamma}}^{\frac{\pi}{\gamma}}
K(v-x)\epsilon(x){\rm d} x
\qquad
\epsilon(v)=\frac{2J\sinh^2 \gamma}{\cos\gamma v-\cosh\gamma}.
\end{equation}
The function $\mfA(v):=1+\mfa(v)$ and $\mfAb(v):=1+\mfab(v)$ are determined
from the following set of non-linear integral equations (NLIEs):
\begin{align}
\ln \mfa(v)&=\left\{-\beta+\lambda_n  (AD)^{n-1}\right\}\varepsilon(v)+
              \K*\ln\mfA(v)-\K*\ln\mfAb(v+2\i-\i\epsilon)\nn \\
\ln \mfab(v)&=\left\{-\beta+\lambda_n  (AD)^{n-1}\right\}
\varepsilon(v)+\K*\ln\mfAb(v)-\K*\ln\mfA(v-2\i+\i\epsilon).
\label{nlie}
\end{align}
Here $\epsilon$ is an infinitesimally small number and
the symbol $\ast$ denotes the convolution $f\ast
g(v)=\int_{-\pi/\gamma}^{\pi/\gamma}f(v-x)g(x) {\rm d} x$.  The kernel $\K(v)$
and the function $\varepsilon(v)$ are given by
\begin{equation}
\K(v)=\frac{\gamma}{2\pi}\sum_{k=-\infty}^{\infty}
  \frac{\e^{-|k|\gamma}}{2\ch k\gamma}\e^{-\i k \gamma v} 
\qquad
\varepsilon(v)=2\pi A K(v)=
2J\sum_{k=-\infty}^{\infty}\frac{\sinh \gamma }{{2\ch k \gamma}}
\e^{-\i k \gamma v}.
\end{equation}
From the above NLIEs \eqref{eigval} and \eqref{nlie}, or just symmetry
arguments, we find that expectation values of the conserved quantities $\langle
\J^{{(2m)}} \rangle_{m\ge 1}$ are always zero: $\langle \J^{{(2m)}}
\rangle_{m\ge 1}=0$.
Therefore, due to the relations \eqref{drude} and \eqref{qtm} (note that
$\J^{(2)}=\J_{\rm th}$), the thermal Drude weight $\D(T)$ is given by
\begin{equation}
\D(T)=\beta^2\bra \J^{(2)2}\ket=
\beta^2 \left(\frac{\partial }{\partial \lambda_2}\right)^2
\ln \Lambda.
\label{drudeqtm}
\end{equation}

We would like to remark on the structure of the NLIEs.  Our NLIEs \eqref{nlie}
are consistent with those for $\lambda_n =0$ in \cite{KlumTH2}, and may be
obtained from those by the replacement of the driving term
\begin{equation}
-\beta \varepsilon(v)\to \left\{-\beta +\lambda_n 
(AD)^{n-1}\right\}\varepsilon(v)
\label{replace}
\end{equation}
reflecting the structure of the general Hamiltonian $\widetilde{\mathcal{H}}$
and the generating function \eqref{generator} (or \eqref{general}).
Employing the same analogy, we can derive an alternative expression based on
the thermodynamic Bethe ansatz (TBA) \cite{Gaudin,Tak,Takabook}.
The resultant TBA equation consist of infinitely many NLIEs.
In contrast to the TBA method, in our approach the thermal quantities are
determined from only two NLIEs, which allows us to evaluate physical
quantities numerically with quite high accuracy.
%
\begin{figure}[ttt]
\begin{center}
\includegraphics[width=0.86\textwidth]{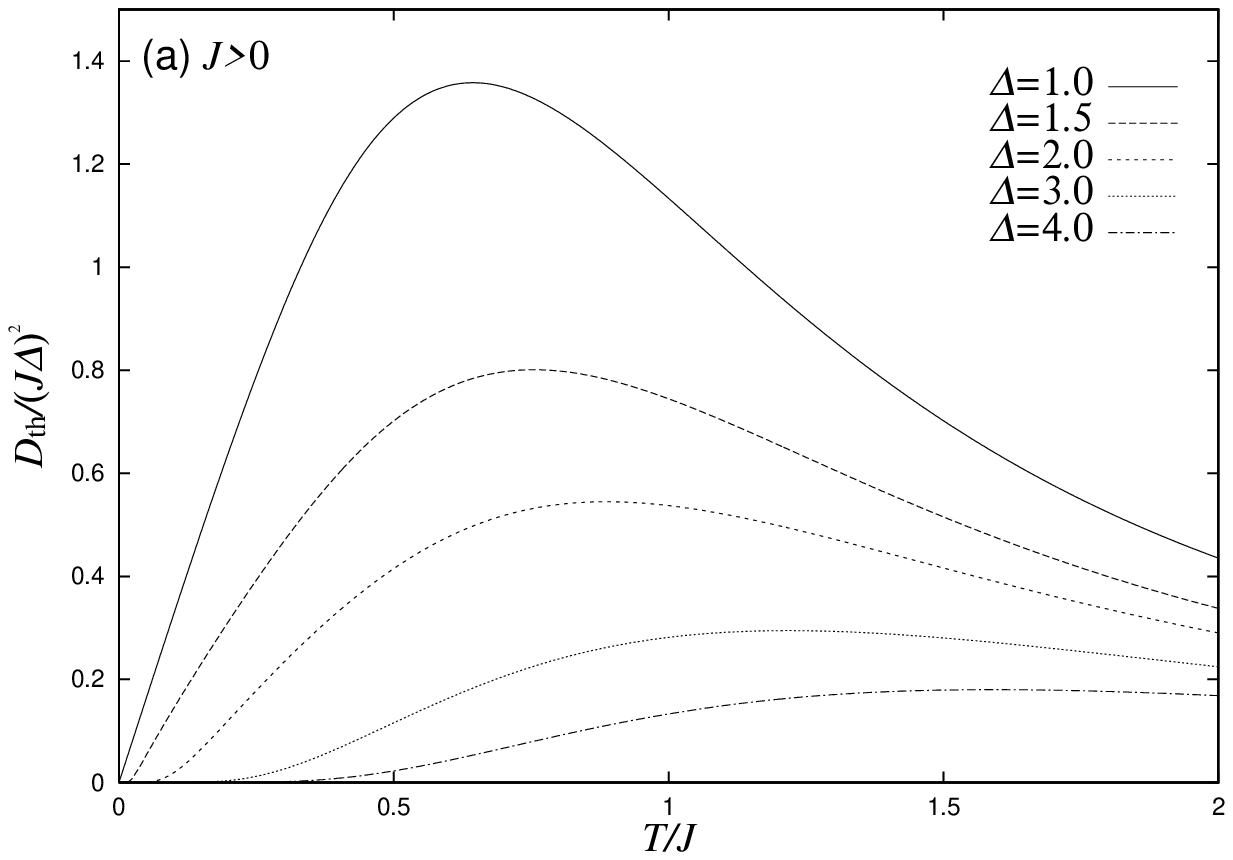}
\includegraphics[width=0.86\textwidth]{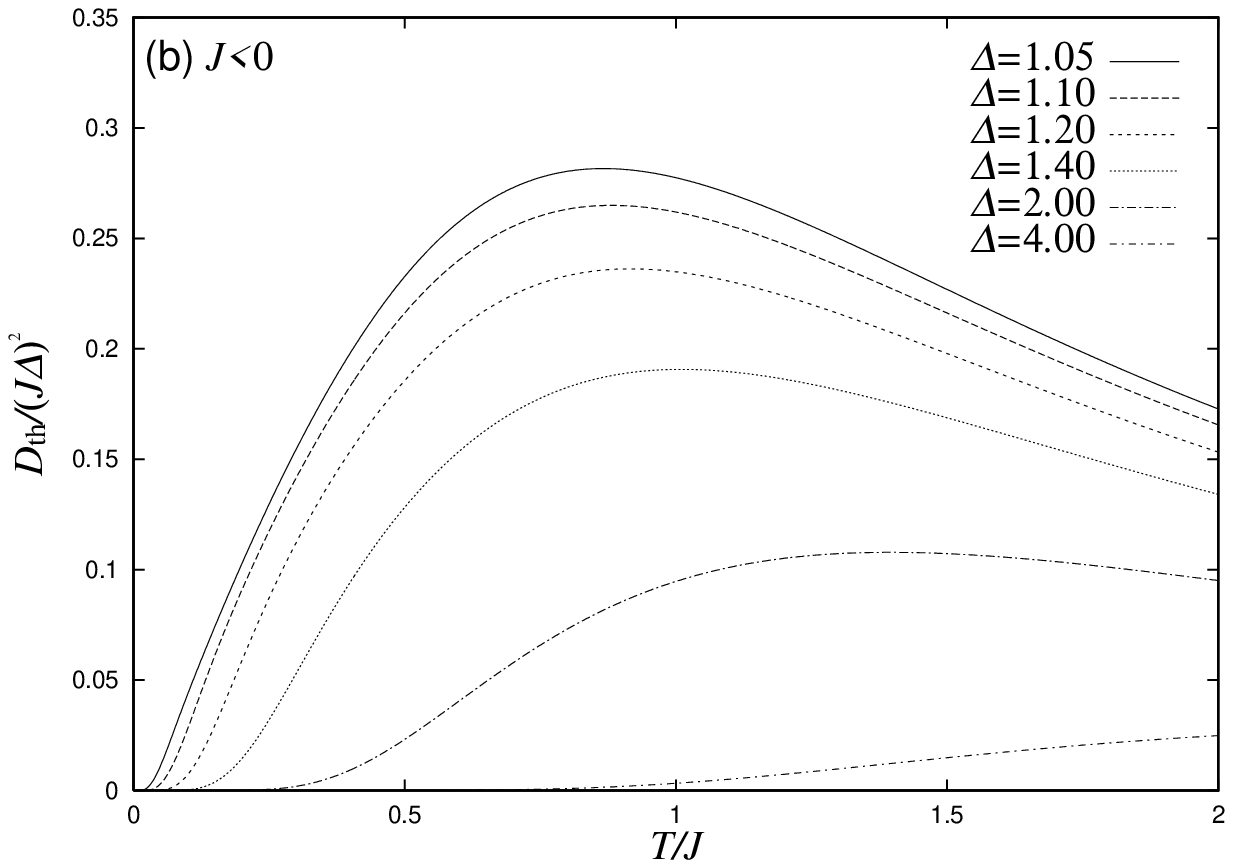}
\end{center}
\caption{Illustration of numerical results for the thermal Drude weight
$D_{\rm th}(T)$ (in units of $(J\Delta)^2$) in the antiferromagnetic $J>0$ (a)
and ferromagnetic $J<0$ (b) regimes as a function of temperature $T$ (in units
of $J$) for various anisotropies.}
\label{f1}
\end{figure}
\begin{figure}[ttt]
\begin{center}
\includegraphics[width=0.86\textwidth]{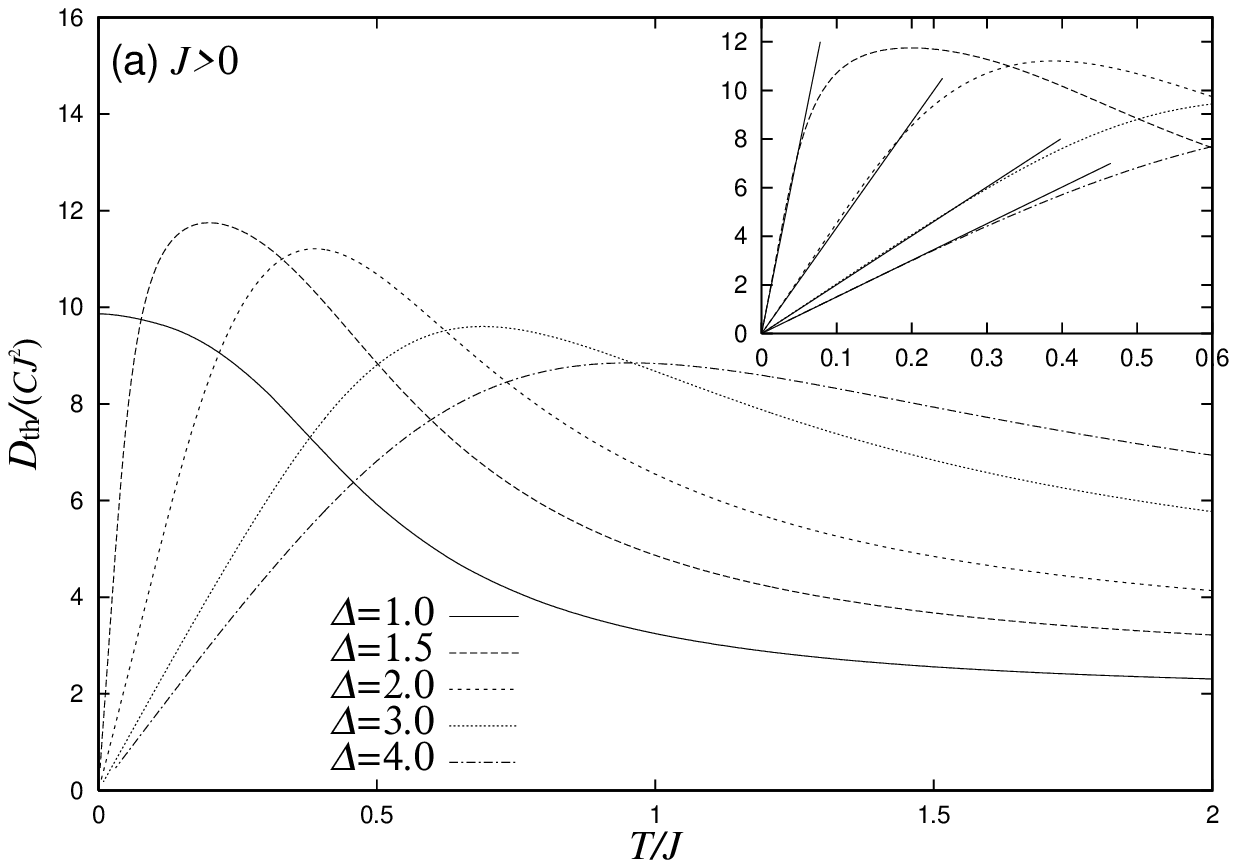}
\includegraphics[width=0.86\textwidth]{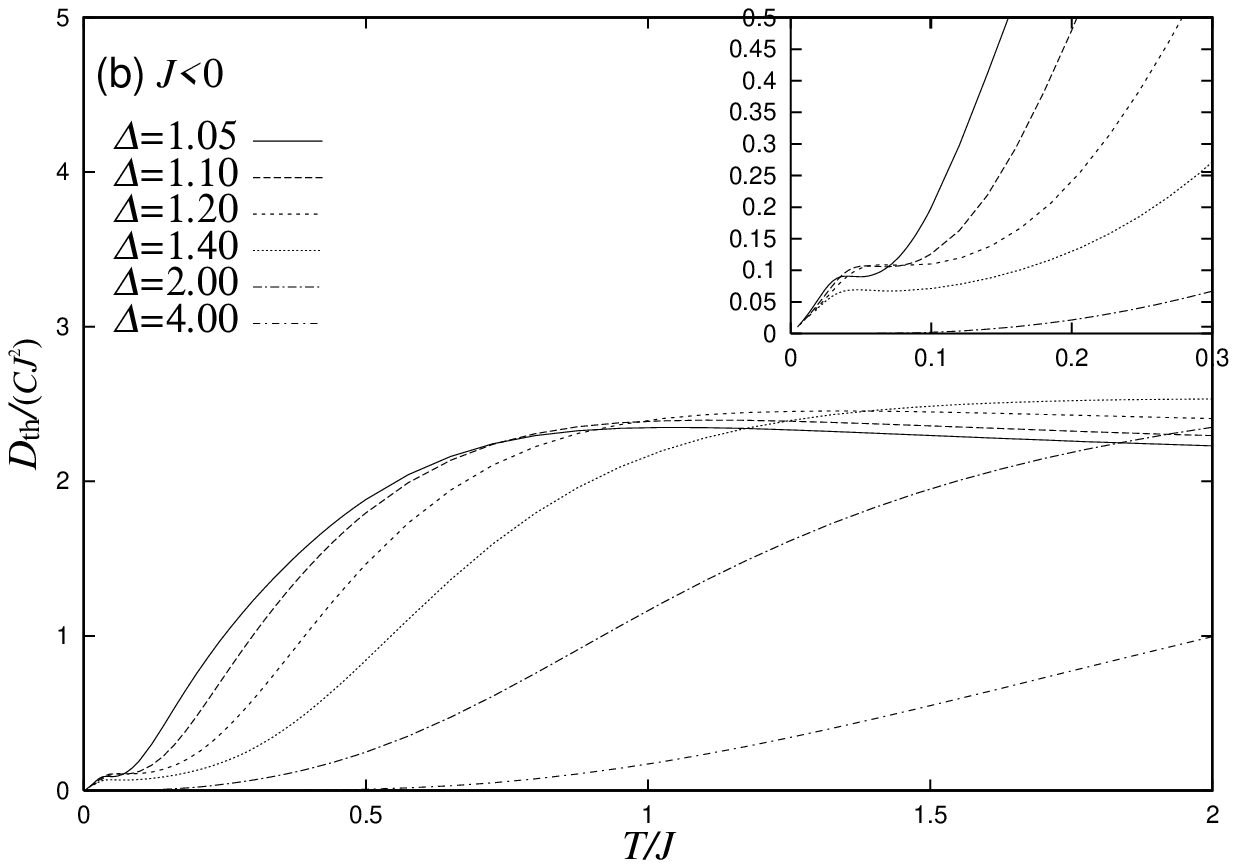}
\end{center}
\caption{Illustration of the ratio of the thermal Drude weight and the
specific heat $\D(T)/C(T)$ (in units of $J^2$) in the antiferromagnetic $J>0$
(a) and ferromagnetic $J<0$ (b) regimes as a function of temperature $T$ (in
units of $J$) for various anisotropies. In the inset numerical data at low
temperatures are shown.  The linear lines in the inset of Fig.\ref{f2}(a) are
the analytical results for the low-temperature asymptotics.}
\label{f2}
\end{figure}
%

In Fig.\ref{f1}, the temperature dependence of the thermal Drude weight
$\D(T)$ is depicted for various anisotropy parameters.  We find that at low
temperatures and $\Delta>1$, the weight $\D(T)$ decays exponentially with
decreasing temperature:
\begin{equation}
\D(T)\sim  \frac{1}{\sqrt{T}}\exp\left(-\frac{\delta}{T}\right)
\quad \text{for $T\ll 1$}.
\end{equation}
where $\delta$ is the energy gap of the one-spinon\footnote{Note that the
excitation gap
(spectral gap) actually appearing in the energy spectrum is the two-spinon gap
$2\delta$.}  (respectively one-magnon) excitation in the antiferromagnetic
(respectively ferromagnetic) regime (see the next section for details).  On
the other hand at high temperatures, $\D(T)$ behaves as
\begin{equation}
\D(T)\sim \frac{1}{T^2} \quad \text{for $T\gg 1$}.
\end{equation}
One finds that $\D(T)$ has a finite temperature maximum and the corresponding
temperature $T_0$ shifts to higher values with increasing interaction strength.
In the Ising limit $\Delta\to\infty$, the temperature $T_0$ moves to infinity
for fixed $J$ or the height of the peak goes to zero for fixed $J\Delta$ 
and then the thermal Drude weight $\D(T)$ is always zero at arbitrary
temperatures.  

In Fig.\ref{f2}, we show the ratio of the thermal Drude weight and the
specific heat $\D(T)/C(T)$ as a function of the temperature $T$ for various
anisotropy parameters. Due to the mass-gap, this ratio goes to zero in the
low-temperature limit (c.f. eq.~(4.12) in \cite{KS} for the massless regime).
In the next section the 
aforementioned high- and low-temperature asymptotics are calculated
analytically.
%
\section{Low- and high-temperature limit}
In this section we consider the low- and high-temperature behavior of the
thermal Drude weight for both ferromagnetic ($J<0$) and antiferromagnetic
($J>0$) regimes.
In general, neither the NLIEs \eqref{nlie} nor the TBA equations, 
can be solved analytically.
However, in the low- and high-temperature limits, the equations simplify
especially for the gapped regimes. Hence we can exactly evaluate the
asymptotic behavior.
%
%
\subsection{Low-temperature limit}
{\bf A. Antiferromagnetic regime ($J>0$)}\\\\
%
Let us consider the low-temperature asymptotics in the antiferromagnetic
regime ($J>0$).
From the NLIEs \eqref{nlie} at low temperatures $\beta\gg 1$, the auxiliary
functions $\mfa(v)$ and $\mfab(v)$ reduce to
\begin{equation}
\mfa(v)=\mfab(v)=\exp\left[
\left\{-\beta+\lambda_n(AD)^{n-1}\right\}\varepsilon(v)\right]
\quad \text{for $\beta\gg1$}.
\label{aprx1}
\end{equation}
The sub-leading terms are exponentially smaller in $T$ than the above
expression.
Substituting \eqref{aprx1} into the relation \eqref{eigval} and using
\eqref{qtm}, we have
\begin{equation}
\bra \mathcal{J}^{(n) 2}\ket-\bra \J^{(n)} \ket^2
=2\int_{-\p}^{\p}K(v)\left\{(AD)^{n-1}\varepsilon(v)\right\}^2
\exp[-\beta \varepsilon(v)]{\rm d} v.
\label{corrh}
\end{equation}
Applying the steepest descent method, we arrive at
\begin{align}
\bra \mathcal{J}^{(n) 2}\ket-\bra \J^{(n)}\ket^2=&
\frac{2(-1)^{n-1}(2J\sh\gamma)^{\frac{4n-1}{2}}}
{\sqrt{-2\pi \alpha_2}}
\e^{\frac{-2 J\alpha_0\sh\gamma}{T}} \nn \\
&\times\left\{
\alpha_0\alpha_{n-1}^2 T^{\frac{1}{2}}+
\frac{\alpha_2\alpha_{n-1}^2+2\alpha_0(\alpha_n^2+\alpha_{n-1}\alpha_{n+1})}
{4J\alpha_2\sh\gamma}T^{\frac{3}{2}}+O(T^{\frac{5}{2}})\right\}
\label{currenthigh}
\end{align}
where
\begin{equation}
\alpha_n=\sum_{k=-\infty}^{\infty}
\frac{(-1)^k k^n}{2\ch k \gamma }.
\end{equation}
Thus by setting $n=2$ and using \eqref{drudeqtm}, we obtain the
low-temperature asymptotics of the Drude weight $\D(T)$
\begin{equation}
\D(T)=\frac{-2(2 J \sinh\gamma)^{\frac{5}{2}}
\alpha_0\alpha_2}{\sqrt{-2\pi\alpha_2}}
\e^{\frac{-2J\alpha_0\sinh\gamma}{T}}
\left\{T^{-\frac{1}{2}}+O\left(T^{\frac{1}{2}}\right)
\right\}.
\label{drudelow1}
\end{equation}
Here we have used the fact $\alpha_{2m-1}=0$ when $m\ge1$. 
Note that the exponent $-2J\alpha_0\sinh\gamma$ in \eqref{drudelow1} is
nothing but the one-spinon excitation energy of the model.
This result indicates that the spinon excitation mainly contributes to
the low-temperature heat conduction in the antiferromagnetic regime.
Comparing the low-temperature limit for the specific heat 
$C(T)=\beta^2(\bra\J^{(1)2}\ket-\bra \J^{(1)}\ket^2)$
\begin{equation}
C(T)=\frac{2 (2J\sinh\gamma)^{\frac{3}{2}}\alpha_0^3}
{\sqrt{-2\pi\alpha_2}}\e^{\frac{-2J\alpha_0\sinh\gamma}{T}}
\left\{T^{-\frac{3}{2}}+\frac{3}{4 J\alpha_0\sinh \gamma}
T^{-\frac{1}{2}}+O\left(T^{\frac{1}{2}}
\right)\right\}
\label{ratio1}
\end{equation}
we have
\begin{equation}
\frac{\D(T)}{C(T)}=\frac{-2 J\alpha_2\sinh\gamma}{\alpha_0^2}T+O(T^2).
\end{equation}
In the inset of Fig.~\ref{f2}(a), we show these results for various anisotropy
parameters.\\\\
%
\noindent
{\bf B. Ferromagnetic regime ($J<0$) }\\\\
%
Next we analyze the low-temperature asymptotics in the ferromagnetic regime
($J<0$). 
In this regime the auxiliary functions $\mfa(v)$ and $\mfab(v)$ are no longer
small at low-temperature, hence the analysis of the NLIEs becomes
difficult. To deal with this situation, we introduce the alternative equations
by taking the reciprocals \cite{KlumTH2}:
\begin{alignat}{2}
&\mfb(v)=\frac{1}{\mfa(v)} &\qquad& 
\mfbb(v)=\frac{1}{\mfab(v)} \nn \\
&\mfB(v)=1+\mfb(v)&\qquad&
\mfBb(v)=1+\mfbb(v).
\end{alignat} 
Applying the Fourier transform, we derive the NLIEs in terms of these
functions:
\begin{align}
\ln \mfb(v)&=\left\{\beta-\lambda_n  (AD)^{n-1}\right\}
\widetilde{\varepsilon}(v)+
              \widetilde{\K}*\ln\mfB(v)-
              \widetilde{\K}*\ln\mfBb(v+2\i-\i\epsilon) \nn \\
\ln \mfbb(v)&=\left\{\beta-\lambda_n  (AD)^{n-1}\right\}
\widetilde{\varepsilon}^{\ast}(v)+\widetilde{\K}*\ln\mfBb(v)-
\widetilde{\K}*\ln\mfB(v-2\i+\i\epsilon)
\label{nlie2}
\end{align}
where the driving terms $\varepsilon(v)$ and $\varepsilon^{\ast}(v)$ are given
by
\begin{equation}
\widetilde{\varepsilon}(v)=  A\gamma 
\frac{\cosh\frac{\gamma}{2}(1-\i v)}
{2\sinh\frac{\gamma}{2}(1-\i v)}
\quad
\widetilde{\varepsilon}^{\ast}(v)= A\gamma 
\frac{\cosh\frac{\gamma}{2}(1+\i v)}
{2\sinh\frac{\gamma}{2}(1+\i v)}
\end{equation}
and the kernel $\widetilde{\kappa}(v)$ is defined as
\begin{equation}
\widetilde{\kappa}(v)=-\frac{\gamma}{2\pi}\sum_{k=1}^{\infty}
\frac{\e^{-k\gamma}\cosh(\i k \gamma v)}
{\sinh{k \gamma}}.
\end{equation}
Accordingly, the largest eigenvalue $\Lambda$ is written as
\begin{align}
&\ln\Lambda=
\int_{-\frac{\pi}{\gamma}}^{\frac{\pi}{\gamma}}\widetilde{K}(v)
\ln\mfB(v){\rm d} v +
\int_{-\frac{\pi}{\gamma}}^{\frac{\pi}{\gamma}}
\widetilde{K}^{\ast}(v)\ln\mfBb(v){\rm d} v \nn \\
&\widetilde{K}(v)=\frac{\widetilde{\varepsilon}(v)}{2\pi A} \qquad
\widetilde{K}^{\ast}(v)=\frac{\widetilde{\varepsilon}^{\ast}(v)}{2\pi A}.
\label{eigval2}
\end{align}

Applying an iteration procedure to \eqref{nlie2}, we obtain the
low-temperature behavior of the auxiliary functions:
\begin{align}
\ln\mfb(v)&=g(v)+\int_{-\frac{\pi}{\gamma}}^{\frac{\pi}{\gamma}}
\widetilde{\kappa}(v-x)e^{g(x)}{\rm d}x-
\int_{-\frac{\pi}{\gamma}}^{\frac{\pi}{\gamma}}
\widetilde{\kappa}(v+2\i-\i\epsilon)e^{g^{\ast}(x)} {\rm d} x\nn \\
\ln\mfbb(v)&=g^{\ast}(v)+\int_{-\frac{\pi}{\gamma}}^{\frac{\pi}{\gamma}}
\widetilde{\kappa}(v-x)e^{g^{\ast}(x)}{\rm d}x-
\int_{-\frac{\pi}{\gamma}}^{\frac{\pi}{\gamma}}
\widetilde{\kappa}(v-2\i+\i\epsilon)e^{g(x)}{\rm d} x
\end{align}
where 
\begin{equation}
g(v)=\left\{\beta-\lambda_n  (AD)^{n-1}\right\}
\widetilde{\varepsilon}(v) \qquad
g^{\ast}(v)=\left\{\beta-\lambda_n  (AD)^{n-1}\right\}
\widetilde{\varepsilon}^{\ast}(v).
\end{equation}
Identifying the poles of $\widetilde{\kappa}(v)$ at $v=\pm 2\i$ and applying
Cauchy's theorem, the dominant contributions of the integrals for $T\ll 1$ are
evaluated
\begin{align}
\int^{\frac{\pi}{\gamma}}_{-\frac{\pi}{\gamma}}\widetilde{\kappa}
(v-x+2\i-\i\epsilon)
\ln(1+\e^{g^{\ast}(x)}) {\rm d}x\sim -\frac{\gamma}{2\pi}
\int^{\frac{\pi}{\gamma}}_{-\frac{\pi}{\gamma}}
\frac{e^{-\frac{\i}{2}\gamma(v-x+\i\epsilon)}}
{2\sinh\frac{\i}{2}\gamma(v-x+\i\epsilon)}
\e^{g^{\ast}(x)}{\rm d}x
=-\e^{g^{\ast}(v)}.
\end{align}
Using this and neglecting the subdominant terms, we obtain
\begin{equation}
\ln\mfB(v)\sim \mfb(v)\sim\e^{g(v)}(1+\e^{g^{\ast}(v)})=
                   \e^{g(v)}+\
                   e^{g(v)+g^{\ast}(v)}.
\end{equation}
For $\ln\mfBb(v)$ a similar equation is valid. Substituting the results
into \eqref{eigval2}, we have
\begin{equation}
\ln\Lambda=\int_{-\frac{\pi}{\gamma}}^{\frac{\pi}{\gamma}}\widetilde{K}(v)
\e^{g(v)}{\rm d} v +
\int_{-\frac{\pi}{\gamma}}^{\frac{\pi}{\gamma}}
\widetilde{K}^{\ast}(v)\e^{g^{\ast}(v)}{\rm d} v+
\int_{-\frac{\pi}{\gamma}}^{\frac{\pi}{\gamma}}
[\widetilde{K}(v)+\widetilde{K}^{\ast}(v)]
\e^{g(v)+g^{\ast}(v)} {\rm d}v.
\label{eigval3}
\end{equation}
We can calculate the first (respectively second) term in \eqref{eigval3} by
shifting the contour to $-\i \infty$ (respectively $\i \infty$).
The third integral is evaluated by a saddle point integration.

Using the relation \eqref{qtm}, we obtain the low-temperature asymptotics of
the conserved quantities:
\begin{align}
\bra \mathcal{J}^{(n) 2}\ket-\bra \J^{(n)}\ket^2=&
(J\sinh\gamma)^2\e^{\beta J \sinh\gamma} \delta_{n1}+
\frac{(-1)^{n-1}(-2J\sh\gamma)^{\frac{4n-1}{2}}}
{\sqrt{-2\pi \beta_2}}
\e^{\frac{2 J\beta_0\sh\gamma}{T}} \nn \\
&\times\left\{
\beta_0\beta_{n-1}^2 T^{\frac{1}{2}}+
\frac{\beta_2\beta_{n-1}^2+2\beta_0(\beta_n^2+\beta_{n-1}\beta_{n+1})}
{-4J\beta_2\sh\gamma}T^{\frac{3}{2}}+O(T^{\frac{5}{2}})\right\}
\label{ferro}
\end{align}
where
\begin{equation}
\beta_n=\sum_{k=-\infty}^{\infty}
(-1)^k k^n\e^{-|k|\gamma}
\quad \beta_0=\frac{\sinh\gamma}{1+\Delta}
\quad \beta_1=0
\quad \beta_2=-\frac{\sinh\gamma}{(1+\Delta)^2}.
\end{equation}
Setting $n=2$, we derive the low-temperature asymptotics of the thermal Drude
weight $\D(T)$:
\begin{equation}
\D(T)=\frac{ (-2 J )^{\frac{5}{2}}(\Delta-1)^2 
\e^{\frac{2J(\Delta-1)}{T}}}{\sqrt{2\pi}}
\left\{T^{-\frac{1}{2}}+O\left(T^{\frac{1}{2}}\right)
\right\}.
\label{drudelow2}
\end{equation}
The exponent $2|J|(\Delta-1)$ in \eqref{drudelow2} agrees with the one-magnon
gap. This result coincides with that from 
numerical diagonalization of the Hamiltonian for size $L=18$ \cite{MHCB}.
For the specific heat $C(T)=\beta^2(\bra \J^{(1)2}\ket-\bra \J^{(1)}\ket^2)$
we obtain
\begin{equation}
C(T)=\frac{(J\sinh\gamma)^2\e^{\frac{J\sinh\gamma}{T}}}{T^2}+
\frac{(-2 J)^{\frac{3}{2}}(\Delta-1)^2 \e^{\frac{2J(\Delta-1)}{T}}}
{\sqrt{2\pi}}\left\{T^{-\frac{3}{2}}+\frac{3}{-4J(\Delta-1)}
T^{-\frac{1}{2}}+O\left(T^{\frac{1}{2}}
\right)\right\}.
\label{specific}
\end{equation}
Here the exponent $|J|\sinh\gamma$ in the specific heat \eqref{specific} is
the one-spinon excitation gap in the ferromagnetic regime. We observe a
crossover behavior from dominant one-magnon excitation ($\Delta>5/3)$ and
dominant one-spinon excitation ($\Delta\le 5/3$) \cite{Takabook,Tak2}.
In contrast to this, a crossover does not take place in the low-temperature
thermal Drude weight, which implies that only the magnon excitation
contributes to the low-temperature heat conduction.  This result yields a
different behavior of the ratio of the thermal Drude weight and the specific
heat \eqref{ratio} for small and large values of $\Delta$, respectively
\begin{equation}
\frac{\D(T)}{C(T)}= \begin{cases}
                    -2 J T +O(T^2),& \text{for $\Delta\le \dfrac{5}{3}$},\\
      \dfrac{ (-2 J )^{\frac{5}{2}}(\Delta-1) 
       \e^{\frac{J\{2(\Delta-1)-\sinh\gamma\}}{T}}}{\sqrt{2\pi}J^2(\Delta+1)}
     \left\{T^{\frac{3}{2}}+O\left(T^{\frac{5}{2}}\right) 
      \right\}, & \text{for $\Delta>\dfrac{5}{3}$}.
\end{cases}            
\label{ratio} 
\end{equation}
The ratio of ${\D(T)}$ and ${C(T)}$ depends linearly on the temperature and is
independent of the anisotropy for $\Delta\le 5/3$.  On the other hand, for
$\Delta>5/3$ the ratio decays exponentially with temperature and explicitly
depends on $\Delta$. This behavior is clearly observed in Fig.\ref{f2}(b).
%
\subsection{High temperature limit}
%
Here we analyze the high-temperature behavior. In this limit the NLIEs
\eqref{nlie} linearize and hence can be analytically solved.  By use of
identities like
\begin{align}
\frac{\partial}{\partial \beta}\ln\mfa&=\frac{\mfA}{\mfa}
\frac{\partial}{\partial \beta}\ln\mfA\\
\frac{\partial^2}{\partial \beta^2}\ln\mfa&=
-\frac{\mfA}{\mfa^2}\left(\frac{\partial}{\partial \beta}\ln\mfA\right)^2
+\frac{\mfA}{\mfa}\frac{\partial^2}{\partial \beta^2}\ln\mfA
\end{align}
and the limiting behavior ${\mfa}=1$, ${\mfA}=2$ for $\beta=0$ we find the
linearized integral equations
\begin{align}
\frac{\partial}{\partial \beta}\ln\mfA(v)
&=
-\frac{1}{2} \varepsilon(v)+
\frac{1}{2}\kappa\ast
\frac{\partial}{\partial\beta}\ln\mfA(v)-
\frac{1}{2}\kappa\ast
\frac{\partial}{\partial\beta}\ln\mfAb(v+2\i-\i\epsilon) \nn \\
\frac{\partial}{\partial \beta}\ln\mfAb(v)
&=
-\frac{1}{2} \varepsilon(v)+
\frac{1}{2}\kappa\ast
\frac{\partial}{\partial\beta}\ln\mfAb(v)-
\frac{1}{2}\kappa\ast
\frac{\partial}{\partial\beta}\ln\mfA(v-2\i+\i\epsilon)
\label{lininbeta}
\end{align}
and integral equations linear in $(\partial/\partial\lambda)^2\ln\mfA$ (with
$\lambda:=\lambda_n$)
\begin{align}
\left(\frac{\partial}{\partial \lambda}\right)^2\ln\mfA(v)
&=
\left(\frac{\partial}{\partial\lambda}\ln\mfA(v)\right)^2+
\frac{1}{2}\kappa\ast
\left(\frac{\partial}{\partial\lambda}\right)^2\ln\mfA(v)-
\frac{1}{2}\kappa\ast
\left(\frac{\partial}{\partial\lambda}\right)^2\ln\mfAb(v+2\i-\i\epsilon) \nn \\
\left(\frac{\partial}{\partial \lambda}\right)^2\ln\mfAb(v)
&=
\left(\frac{\partial}{\partial\lambda}\ln\mfAb(v)\right)^2+
\frac{1}{2}\kappa\ast
\left(\frac{\partial}{\partial\lambda}\right)^2\ln\mfAb(v)-
\frac{1}{2}\kappa\ast
\left(\frac{\partial}{\partial\lambda}\right)^2\ln\mfA(v-2\i+\i\epsilon).
\end{align}
The current correlator \eqref{currentcorr} is found from \eqref{qtm}
\begin{align}
\left(\frac{\partial}{\partial \lambda}\right)^2\ln\Lambda
\biggr|_{\lambda=0}&=
\int_{-\frac{\pi}{\gamma}}^{\frac{\pi}{\gamma}}K(v)
\left(\frac{\partial}{\partial \lambda}\right)^2\ln\mfA(v)\mfAb(v)
\biggr|_{\lambda=0}{\rm d}v \nn \\
&=-\frac{1}{\pi A}\int_{-\frac{\pi}{\gamma}}^{\frac{\pi}{\gamma}}
\frac{\partial}{\partial\beta}\ln\mfA(v)\left(
\frac{\partial}{\partial\lambda}\ln\mfA(v)\right)^2
\biggr|_{\lambda=0}{\rm d}v +
{\rm c.c.}.
\end{align}
where in the last line we have used the dressed function formalism.  With the
high temperature asymptotics $\mfa(v), \mfab(v)= 1$ for $\beta=0$ we find
\begin{equation}
\left(\frac{\partial}{\partial \lambda}\right)^2\ln\Lambda
\biggr|_{\lambda=0}
=-\frac{1}{8\pi A}\int_{-\frac{\pi}{\gamma}}^
{\frac{\pi}{\gamma}}
\frac{\partial\mfa(v)}{\partial\beta}\left(
\frac{\partial\mfa(v)}{\partial\lambda}\right)^2\biggr|_{\lambda=0}{\rm d}v 
+{\rm c.c.}.
\end{equation}
The integrands in the above equation are found analytically. First, from
\eqref{lininbeta} we obtain
\begin{align}
\frac{\partial\mfa(v)}{\partial \beta}\biggr|_{\lambda=0}&=
\frac{J\sinh^2\gamma}{2\sin\frac{\gamma}{2}(v+\i)}
\left(\frac{1}{\sin\frac{\gamma}{2}(v+3\i)}-
      \frac{1}{\sin\frac{\gamma}{2}(v-\i)}\right)
\end{align}
and $\mfab$ is the complex conjugate.

Similar to the reasoning at the beginning of this section we obtain linear
integral equations for the derivatives with respect to $\lambda$
\begin{align}
\frac{\partial}{\partial \lambda}\ln\mfA(v)
&=
\frac {(AD)^{n-1}}{2} \varepsilon(v)+
\frac{1}{2}\kappa\ast
\frac{\partial}{\partial\lambda}\ln\mfA(v)-
\frac{1}{2}\kappa\ast
\frac{\partial}{\partial\lambda}\ln\mfAb(v+2\i)\nn \\
\frac{\partial}{\partial \lambda}\ln\mfAb(v)
&=
\frac {(AD)^{n-1}}{2}\varepsilon(v)+
\frac{1}{2}\kappa\ast
\frac{\partial}{\partial\lambda}\ln\mfAb(v)-
\frac{1}{2}\kappa\ast
\frac{\partial}{\partial\lambda}\ln\mfA(v-2\i).
\end{align}
Hence we find in the high temperature limit
\begin{align}
\frac{\partial}{\partial \lambda}\ln\mfA(v)\biggr|_{\lambda=0}&=
-(AD)^{n-1}\frac{\partial}{\partial \beta}\ln\mfA(v)\biggr|_{\lambda=0}\\
\frac{\partial}{\partial \lambda}\mfa(v)\biggr|_{\lambda=0}&=
-(AD)^{n-1}\frac{\partial}{\partial \beta}\mfa(v)\biggr|_{\lambda=0}.
\end{align}
By use of these explicit expressions for $n=2$ we obtain the high temperature
limit of $\D(T)$
\begin{equation}
\D(T)=\frac{ J^4(2+\cosh2\gamma)}{2}\frac{1}{T^2}+
O\left(\frac{1}{T^3}\right)
\end{equation}
which is consistent with the result in \cite{KS}\footnote{There is a misprint
in (4.18) in \cite{KS}: on the r.h.s of (4.18) a factor $\pi$ is missing.}
after changing the parameter $\gamma\to\i \gamma$.
%
\section{Summary and discussion}
%
In this paper we have discussed the thermal transport properties in the
massive regime for the spin-1/2 $XXZ$ chain.
The thermal current operator is expressed as a conserved quantity resulting
into an anomalous thermal transport.  The thermal Drude weight at finite
temperatures was calculated by a path integral formulation. Due to finite
temperature effects, non-dissipative thermal transport was observed even in
the massive regime of the model.
At low-temperatures ($T\ll 1$), $\D(T)$ can be written in a universal form:
$\D(T)\sim\e^{-\delta/T}/\sqrt{T}$, where $\delta$ is the one-spinon
(respectively one-magnon) gap for the antiferromagnetic (respectively
ferromagnetic) regime.

Finally, we would like to remark some  generalizations of our 
results.  (i) As mentioned in Sec.~2, the conservation law 
of the thermal current is not limited to the present model. 
Therefore the present approach is directly applicable to more general models
such as the $XYZ$ and the integrable higher-spin or higher-rank chains.  (ii)
It is quite interesting to consider the system in an external magnetic field
$h$:
$H=\mathcal{H}+h\sum_j \sigma_j^z/2$.
From the definition \eqref{lc}, it is observed that the energy current
$\J_{\rm E}$ includes the spin current $\J_{\rm s}$ proportional to
$h$. Namely
\begin{equation}
\J_{\rm E}=\J_{\rm th}+h \J_{\rm s}\qquad
\J_{\rm s}=\i J\sum _{j=1}^{L}(\sigma_j^{+}\sigma_{j+1}^{-}-
\sigma_{j+1}^{+}\sigma_{j}^{-})
\end{equation}
where the explicit form of $J_{\rm th}$ is given by \eqref{tc}.
Despite the fact that the thermal current is still conserved $[H,\J_{\rm
th}]=0$, the energy current is no longer a conserved quantity because
$[H,\J_{\rm s}]\ne 0$.
To evaluate the thermal conductivity at $h>0$,
we should take carefully into account thermomagnetic effects, 
which does not matter when $h=0$ (being equivalent to half-filling) because
$\bra \J_{\rm s}\J_{\rm th} \ket=0$.
The correct thermal Drude weight at $h>0$ is determined from linear-response
theory (see \cite{Mahan} for example):
\begin{equation}
\D(T,h)=\beta^2\bra \J_{\rm th}^2\ket_h- \frac{1}{\beta} 
S(T,h)^2 D_{\rm s}(T,h) \qquad
S(T,h)=\beta^2 \frac{\bra \J_{\rm s}\J_{\rm th}\ket_{h}}
{D_{\rm s}(T,h)}
\label{drudeh}
\end{equation}
where $\bra\cdots\ket_h$ denotes the thermal expectation value per site
for the system
$H$; $S(T,h)$ is the thermomagnetic power and $D_{\rm s}(T,h)$ is the Drude
weight of the spin stiffness $\sigma(\omega)$:
\begin{align}
&\Re\sigma(\omega)=\pi D_{\rm s}(T,h)
\delta(\omega)+\sigma_{\rm reg}(\omega) \nn \\
&D_{\rm s}(T,h)=\frac{1}{L}\left\{\bra -K\ket_{ h}
-2\sum_{ E_n\ne E_m}p_n \frac{
\bigl|\langle n |\J_{\rm s} |m\rangle \bigr|^2}
 {\E_m-\E_n}\right\} \quad p_n=\frac{\e^{-\beta \E_n}}
{\sum_k \e^{-\beta \E_k}}.
\end{align}
Here, $K$ and $\E_n$ denote the kinetic energy and the energy eigenvalues of
the Hamiltonian $H$, respectively.
The first term in \eqref{drudeh} can be evaluated by an extension of the
present method.  The Drude weight $D_{\rm s}(T,h)$ in the second term may be
calculated by the approach given in \cite{FK98}. In fact in the massless
regime ($\Delta\le 0$) with zero external field, $D_{\rm s}(T,0)$ has been
already calculated by Zotos \cite{Zotos99}.
As mentioned in the preceding section, however, the validity of the resultant
Drude weight is presently debated (see \cite{ZP03,NarMA98,AlGros1,FK03} for
example).  In this respect, a rigorous study 
of the Drude weight together
with the thermomagnetic power is highly desired to investigate the thermal
transport in finite magnetic fields.
%
\section*{Acknowledgments}
We would like to thank M. Shiroishi, M. Takahashi and J. Takeya
for stimulating discussions. The authors acknowledge financial 
support by the Deutsche Forschungsgemeinschaft under grant 
No.~Kl 645/4-1 and SP1073.
KS is supported in part by the JSPS research fellowships for 
young scientists.
%
%
%
%

%

\end{document}